\documentclass[seceq]{ptptex}
\preprintnumber[3cm]{YITP-10-88}
\title{%        %You can use \\ for explicit line-break.
Pseudoscalar mesons in nuclei and 
partial restoration of chiral symmetry
}

\author{%       %Use \scshape for the family name.
Daisuke \textsc{Jido}%
}

\inst{%     %Affiliation, neglected when [addenda] or [errata].
Yukawa Institute for Theoretical Physics, Kyoto University, Kyoto 606-8502, Japan
}

\abst{%         %This abstract is neglected when [addenda] or [errata].
Recent theoretical developments of hadrons in nuclei are briefly 
reviewed in the aspect of chiral symmetry in nuclei. We show a
general sum rule connecting in-medium hadronic quantities and quark
condensate, emphasizing that both pionic mode and particle-hole
excitations in the nuclear medium are responsible for reduction 
of the chiral condensate in nuclear matter. We also discuss that 
few-body nuclear systems with kaons are good objects 
to investigate basic features of kaon-nucleus interactions,
showing recent understandings of the structure of $\Lambda(1405)$
and possible bound states of $\bar KNN$ and $K\bar KN$. 
We also mention eta mesonic nuclei in connection with chiral symmetry
of baryons.  
}

\begin{document}

\maketitle

%%%%%%%%%%%%%%%%%%%%%%%%%%%
\section{Introdution}
%%%%%%%%%%%%%%%%%%%%%%%%%%%
One of the goals of the contemporary nuclear physics 
is to understand the QCD vacuum structure at 
finite density and/or temperature, especially the fate of
dynamical (or spontaneous) breaking of chiral symmetry.
The broken chiral symmetry at zero density and 
temperature is believed to be restored in certain high 
densities and temperatures. Such restoration of 
chiral symmetry may partially take place even in  
lower densities, such as nuclear density, with effective 
reduction of the quark condensate from 
the in-vacuum value. 
There are a lot of theoretical attempts being done to
reveal the QCD phase structure. In phenomenological
point of view, it is extremely necessary to make a connection 
of such theoretical investigations to experimental observations
by confirming the probable partial restoration of chiral symmetry 
by hadronic experimental observations,
and by finding out the density dependence of the quark condensate. 
One of the promising approaches to make the connection
of finite density QCD with experimental observations is to investigate 
modification of hadron properties in nuclei, since nuclei 
provide us observable finite density systems in which 
one can create or inject hadrons by experiments.
Thus, hadrons in nuclei are good research objects to
investigate in-medium properties of hadrons both
theoretically and experimentally.
Especially, the Nambu-Goldstone bosons, $\pi$, $K$
and $\eta$ mesons, can be good probes to investigate 
chiral symmetry in nuclear medium.

What we can observe for meson properties in experiments
are bound state structures of meson-nucleus systems,
meson decay (or absorption) properties, and low-energy
scattering of meson and nucleus. Recently, precise 
measurements of the level structure of deeply bound 
pionic atoms were performed~\cite{deeppiexp,Suzuki:2002ae}, 
and detailed determination 
of the pion optical potential parameters was carried out. 
Especially, the parameter for the repulsive enhancement of 
the isovector $\pi^{-}$-nucleus interaction was accurately 
extracted as $b_{1}^{\rm free}/b_{1} = 0.78 \pm 0.05$
at around $\rho\sim 0.6 \rho$~\cite{Suzuki:2002ae}. 
Since the $b_{1}$ parameter
is related to the in-medium pion decay constant, if one
assumes the in-medium Weinberg-Tomozawa 
relation~\cite{Kolomeitsev:2002gc,Jido:2008bk}, 
the experimental finding of the $b_{1}$ enhancement 
is expected to be a signal of the reduction of the pion 
decay constant in nuclear matter. The $b_{1}$ 
repulsive enhancement was also seen in low-energy
pion-nucleus scatterings~\cite{Friedman:2004jh}.
These determinations of the pion optical potential parameters
are useful also for giving basic constraints for
pion condensation in neutron star\cite{Ohnishi:2008ng}.
%It is a very important theoretical issue to connect 
%the experimentally observed pion quantities with
%possible partial restoration of chiral symmetry 
%in the nuclear medium.

%%%%%%%%%%%%%%%%%%%%%%%%%%%
\section{Quark condensate in nuclear medium and in-medium pion properies}
%%%%%%%%%%%%%%%%%%%%%%%%%%%
A recently analysis showed an exact relation 
connecting the in-medium quark condensate to hadronic quantities 
in symmetric nuclear matter at the chiral limit~\cite{Jido:2008bk}:
\begin{equation}
   \sum_{\alpha} {\rm Re}\left[ (N_{\alpha}^{*} + F_{\alpha}^{*}) G_{\alpha}^{*1/2} \right]
       = - \langle \bar q q \rangle^{*} \ , \label{eq:exactSR}
\end{equation}
where the in-medium quark condensate is given by 
$\langle \bar q q \rangle^{*} \equiv \langle \Omega | \frac{1}{2} \bar \psi \psi | \Omega 
\rangle$ with the quark field $\psi^{T}=(u,d)$ and the isospin symmetric
matter ground state $|\Omega\rangle$ normalized by 
$\langle \Omega | \Omega \rangle =1$, and the summation is taken over 
all of the zero modes $\alpha$ with the propeties $\varepsilon_{\alpha} \to 0$
as $\vec k_{\alpha} \to \vec 0$ in nuclear matter. 
In Eq.~(\ref{eq:exactSR}), hadronic quantities $N_{\alpha}^{*}$, 
$F_{\alpha}^{*}$ and $G_{\alpha}^{*1/2}$ are matrix elements 
of the axial current $A_{\mu}^{a}$ and the pseudoscalar density 
$\phi_{5}^{a}\equiv \bar \psi i \gamma_{5} (\tau^{a}/2) \psi$ with 
the Pauli matrix $\tau^{a}$ in the flavor space, 
defined by
\begin{eqnarray}
   \langle \Omega_{\ell}^{b}(k)| \phi_{5}^{a}(x) | \Omega \rangle &=&
      \delta^{ab} G_{\ell}^{*1/2} e^{ik\cdot x} \ , \\
   \langle \Omega| A_{\mu}^{a}(x) | \Omega_{\ell}^{b}(k) \rangle &=&
      i\delta^{ab} [n_{\mu} (n\cdot k) N_{\ell}^{*} + k_{\mu} F_{\ell}^{*}] e^{ik\cdot x}    .
\end{eqnarray}
where $| \Omega_{\ell}^{a} \rangle $ with the isospin label $(a=1,2,3)$ 
are the eigenstates of the QCD Hamiltonian 
normalized by $\langle \Omega_{\ell}^{a} | \Omega_{\ell^{\prime}}^{b} \rangle 
= \delta^{ab} 2 \varepsilon_{\ell} (2\pi)^{3} V \delta_{\vec p, \vec p^{\, \prime}}$
with the eigenvalue $\varepsilon_{\ell}$ measured from the ground state and 
spacial volume $V$, and the Lorentz four-vector $n_{\mu}$ specifies 
the frame of the nuclear matter. In the nuclear matter rest frame, 
$n_{\mu}=(1,0,0,0)$, one empirically introduces the temporal and spacial 
``decay constant" as
\begin{equation}
   \langle \Omega | A_{0}^{a}(0) | \Omega_{\alpha}^{b} \rangle =
    i \delta^{ab} \varepsilon_{\alpha} F_{\alpha}^{t}\ , \quad 
   \langle \Omega | A_{i}^{a}(0) | \Omega_{\alpha}^{b} \rangle =
    i \delta^{ab} k_{i\alpha} F_{\alpha}^{s}    \ ,
\end{equation}
which relate to $N_{\alpha}^{*}$ and $F_{\alpha}^{*}$ as
\begin{equation}
   F_{\alpha}^{t} =  N_{\alpha}^{*} + F_{\alpha}^{*} \ , \quad 
   F_{\alpha}^{s} = N_{\alpha}^{*} \ .
\end{equation}

The sum rule shown in Eq.~(\ref{eq:exactSR}) was derived using the 
operator relations of the axial current and the pseudoscalar density 
based on chiral symmetry\footnote{Details of the derivation are given 
in Ref.~\citen{Jido:2008bk}. Extension of the sum rule with finite quark masses
is also discussed there.}.
Consequently, the sum rule is applicable 
for any hadronic density. In vacuum, the sum rule is reduced to 
the well-know Glashow-Weinberg relation~\cite{Glashow:1967rx} 
$F_{\pi}G_{\pi}^{1/2} = - \langle \bar qq \rangle$. One of the 
important consequences is that one has to sum up all the zero
modes in nuclear matter to obtain the in-medium quark condensate. 
Because, in general, the pion mode is not only the zero mode in nuclear matter 
and particle-hole excitations also can be zero modes, the particle-hole excitations 
are also responsible for the in-medium modification of the quark condensate. 
Thus, in order to conclude reduction of the quark condensate 
in nuclear medium, one does not have to separate
the pion mode from nuclear many-body dynamics. 
For further quantitative discussion, one needs dynamical description 
of pion in nuclear matter for actual theoretical calculations of the 
matrix elements. Alternatively, once we extract the 
matrix elements $N_{\alpha}^{*}$ and $F_{\alpha}^{*}$ from 
experimental observables, the sum rule is available for experimental
confirmation of partial restoration of chiral symmetry in nuclear medium.

The exact sum rule can be simplified at the low density. Since the 
one $p$-$h$ excitation decouples from the sum rule 
and higher particle-hole excitations contribute beyond linear density,
only the pion mode saturates the sum rule at linear density:
\begin{equation}
   F^{t}_{\pi} G_{\pi}^{*1/2} = - \langle \bar qq \rangle^{*}  \label{eq:GW}
\end{equation}
where $F^{t}_{\pi}$ is the temporal component of the pion decay 
constant in nuclear medium and $G_{\pi}^{*1/2}$ represents in-medium
wavefunction normalization. By taking a ratio of Eq.~(\ref{eq:GW})
and in-vacuum Glashow-Weinberg relation, we obtain a scaling law valid only for linear density:
\begin{equation}
    \left(\frac{F_{\pi}^{t}}{F_{\pi}}\right) Z^{*1/2}_{\pi} = 
    \frac{\langle \bar qq \rangle^{*}}{\langle \bar qq \rangle} \ ,
    \label{eq:ScaleLaw}
\end{equation}
where $Z_{\pi}^{*} \equiv G_{\pi}^{*}/G_{\pi}$ is the in-medium wavefunction 
renormalization.  Thus, the in-medium modification of the quark condensate
is represented by the in-medium change of the pion decay constant and 
the pion wavefunction renormalization. The importance of the wavefunction 
renormalization in in-medium chiral effective theory is also emphasized 
in Refs.~\citen{Jido:2000bw,Kolomeitsev:2002gc}. 
The in-medium modification of the pion decay constant $F_{\pi}^{t}/F_{\pi}$
can be related to the repulsive enhancement of the $s$-wave 
isovector $\pi$-nucleus interaction observed in deeply bound pionic atoms
and low-energy pion-nucleus scatterings using the in-medium Weinberg-Tomozawa
relation~\cite{Kolomeitsev:2002gc,Jido:2008bk}. The wavefunction 
renormalization at low density can be obtained by 
$\pi$-nucleon scattering~\cite{Jido:2008bk}. Combining these two obserbation
and using the low-density scaling law~(\ref{eq:ScaleLaw}), 
reduction of the quark condensate in nuclei was confirmed
in Ref.~\citen{Jido:2008bk}.
It would be very interesting,
if we could observe the wavefunction renormalization $Z^{*1/2}_{\pi}$
directly in pion-nucleus systems, since it enables us to conclude 
the reduction of the quark condensate from observations of pion-nucleus
systems. It is also important to extend the scaling law (\ref{eq:ScaleLaw})
to higher density. 

%%%%%%%%%%%%%%%%%%%%%%%%%%%
\section{Few-body systems with kaons}
%%%%%%%%%%%%%%%%%%%%%%%%%%%

Kaon is one of the Nambu-Goldstone boson associated with spontaneous 
breaking of chiral symmetry. Although the connection of the in-medium 
properties of kaon to the fate of the SU(3)$_{L}\otimes$SU(3)$_{R}$ 
chiral symmetry in nuclear medium is not clear due to the heavy strange 
quark mass, in-medium properties of kaons are very interesting issues 
both theoretically and experimentally, giving fundamental information of 
kaon condensation in highly dense matter. 
Since $\bar KN$ interaction is strongly attractive,  $\bar K$ may be bound
in nucleus, but the width of the bound states can be large due to strong 
absorption of $\bar K$ into nucleons. This provides difficulties 
for direct experimental observations of $\bar K$ bound states in nuclei.  
To understand fundamental $\bar KN$ interaction, it may be better to 
investigate simpler kaonic nuclear systems first. 

A recent study~\cite{Hyodo:2008xr}
has confirmed within the chiral unitary framework that the $\Lambda(1405)$ 
resonance can be a quasibound state of $\bar KN$ decaying to 
$\pi\Sigma$ with strong interaction. Thus, the $\Lambda(1405)$ resonance 
can be an elementary object of kaon in nucleus. The structure of 
the $\Lambda(1405)$ resonance is so interesting that 
the $\Lambda(1405)$ is composed by two states which have 
different coupling nature to meson-baryon state~\cite{Jido:2003cb}. 
According to chiral dynamics, the state relevant for the $\bar KN$ 
interaction is located around 1420 MeV instead of the nominal 
position 1405 MeV, having a dominant coupling to $\bar KN$.
Thus, it is very important to pin down experimentally the $\Lambda(1405)$ 
resonance position observed in $\bar KN \to \pi \Sigma$.
One of the promising reactions for this purpose is $K^{-}d \to \pi\Sigma n$,
in which the $\Lambda(1405)$ is produced by the $\bar KN$ channel,
as discussed in Ref.~\citen{Jido:2009jf}. In fact, this process was 
already observed in an old bubble chamber experiment~\cite{Braun:1977wd}
and the resonance position was found around 1420~MeV.
Further experimental investigation will be performed in forthcoming experiments 
at J-PARC~\cite{Noumi:JPARC} and DAFNE~\cite{Jido:2010rx}. 

Developing the idea of the $\Lambda(1405)$ as a two-body $\bar KN$ 
quasibound further, we discuss three-body systems with kaons. 
Since a possible $\bar K NN$ quasibound state was theoretically predicted
by Refs.~\citen{nogami,akaishi02}, various approaches have been applied 
to this system~\cite{Ikeda:2007nz,shevchenko07,dote08,Wycech:2008wf}
for calculations of the binding energy and width. 
The present theoretical achievement is that the $\bar KNN$ is bound 
with a large width, but there is controversy over the detailed values 
of the binding energy and width. One of the important issues for the $\bar KNN$
system is whether the possible bound state could be described essentially
by a $\bar KNN$ single channel or some coupled channel effects of
$\pi\Sigma N$ should be implemented to describe the nature of 
the $\bar KNN$ bound state. 
The $\pi\Sigma$ dynamics is also relevant to the structure of the $\Lambda(1405)$,
especially its lower state located around 1390 MeV with 100 MeV 
width~\cite{Jido:2003cb}. Due to lack of experimental information of the 
$\pi\Sigma$ interaction, the pole position of the lower $\Lambda(1405)$ state 
is not precisely determined yet~\cite{Hyodo:2007jq}. 
Detailed experimental information of the $\pi\Sigma$ dynamics,
such as the scattering length and effective range of the $\pi\Sigma$ with
$I=0$~\cite{Ikeda:2010}, is favorable for further understanding of the 
$\Lambda(1405)$ and $\bar KNN$ system. 

Recently, another kaonic three-body system, $K\bar KN$, was investigated
in a nonrelativistic potential model and a possible quasibound state with
$I=1/2$ and $J^{P}=1/2^{+}$ ($N^{*}$) was found with a mass 1910 MeV and a 
width 90 MeV~\cite{Jido:2008zz}. Later this system was studied also in 
a three-body Faddeev approach with coupled channels and a very 
similar state was found~\cite{MartinezTorres:2008kh}.
It was also found in the Faddeev approach~\cite{MartinezTorres:2010zv} 
that this quasibound state is essentially described by a $K\bar K N$ single 
channel, in which $\Lambda(1405)$ and $a_{0}(980)$ are formed in subsystems 
of $\bar KN$ and $\bar KK$, respectively~\cite{Jido:2008zz}. 
This quasibound state is a loosely bound
system having a 20 MeV binding energy. It is found in Ref.~\citen{Jido:2008zz}
that the quasibound has a spatially larger size than typical baryon resonances
by showing  
that the root mean squared radius is as large as 1.7 fm and 
the inter-hadron distances are comparable with 
nucleon-nucleon distances in nuclei. The main decay modes of this state
may be three-body decays, such as $\pi\Sigma K$ and $\pi \eta N$
from the decays of the $\Lambda(1405)$ and $a_{0}(980)$ subsystems. 
The nucleon resonance with
1910 MeV mass and $J^{P}=1/2^{+}$ proposed in 
Refs.~\citen{Jido:2008zz,MartinezTorres:2008kh,MartinezTorres:2010zv}
as a quasibound state of $K\bar KN$ is not listed in the Particle data table 
yet. Experimentally this state could be observed in $\gamma p \to K^{+} \Lambda$.
Further discussion on experimental observation of this $N^{*}$ can be 
found in Ref.~\citen{MartinezTorres:2009cw}.

The quasibound states of $\bar KNN$ and $K\bar K N$ are analogous states
in a sense that real kaons are the constituents in the systems and 
their subsystems, $\bar KN$ and $\bar KK$, have quasibound states with
about 10 MeV binding energy. 
It would be very interesting if we could observe both states experimentally.
Together with further detailed information of the positions of the
double poles of $\Lambda(1405)$, experimental observations 
of the three-body bound states will give us fundamental knowledge of $\bar KN$
and $\bar K$-nucleus interactions. 

%%%%%%%%%%%%%%%%%%%%%%%%%%%
\section{$\eta$ mesonic nuclei}
%%%%%%%%%%%%%%%%%%%%%%%%%%%
The $\eta$ meson is also one of the Nambu-Goldstone bosons. 
It has a neutral charge and interacts with nuclei by strong force. 
Possible bound states of $\eta$ in nuclei were first predicted in Ref.~\citen{Haider},
which was based on the attractive $\eta N$ interaction. 
One of the interesting points in physics of $\eta$ mesonic nuclei
is that one could probe chiral symmetry for baryon with the formation 
spectra of the $\eta$ mesonic nuclei~\cite{Jido:2002yb}
since the $\eta N$ system strongly couples to the $N(1535)$ nucleon 
resonance and $N(1535)$ is a candidate of the chiral partner of 
nucleon~\cite{DK,cdmod}. If nucleon and $N(1535)$ are chiral partners,
their mass difference is expected to be reduced as chiral symmetry 
is partially restored in nuclei. This reduction of the mass difference
induces level crossing between the eta and $N^{*}$-hole modes in 
nuclei and it can be seen in the spectrum shape of 
the formation cross section of $\eta$ mesonic nuclei~\cite{Jido:2008ng}.
For experimental formation of $\eta$ mesonic nuclei, recoil-free reactions 
are essential to produce bound states selectively~\cite{Hayano:1998sy,Nagahiro:2003iv}.
Further details of the formation spectra are discussed in Ref.~\citen{Nagahiro:2003iv}.

%%%%%%%%%%%%%%%%%%%%%%%%%%%
\section{Summary}
%%%%%%%%%%%%%%%%%%%%%%%%%%%

One of the promising ways to learn basic properties of the QCD vacuum 
structure at low density is that one produces hadrons in nuclei experimentally 
and investigates their properties. Deeply bound pionic atoms and low-energy 
pion-nucleus scatterings are most successful systems to complete our 
story connecting experimental observations with partial restoration 
of chiral symmetry in nucleus. This achievement was based on the 
recent theoretical finding of the exact sum rule showing connection 
between the in-medium hadronic quantities and quark condnesate. 
This sum rule tells us that both pionic mode and particle-hole excitations
contribute to in-medium quark condensate, and can be simplified
in low density into the scaling law in which the in-medium change of 
the quark condensate is expressed by the in-medium modification of
the pion decay constant and pion wavefunction renormalization. 
Investigation of mesons in nuclei is closely related to physics of 
baryon resonances in nuclei. For kaonic nuclei, one has to understand 
in-medium properties of $\Lambda(1405)$. For this purpose, as a first step,
few-body nuclear systems with kaons, such as $\bar KNN$ and $K \bar KN$,
are good to be investigated both theoretically and experimentally.
For eta mesonic nuclei, the $N(1535)$ resonance may play an important
role and one could probe chiral symmetry for baryons. 

\section*{Acknowledgements}
The auther would like to thank T.~Hatsuda, T.~Kunihiro,
Y.~Kanada-En'yo, S.~Hirenzaki, H.~Nagahiro, T.~Hyodo
A.M.~Torres and Y.~Ikeda for their collaborations. 
This work was supported by the Grant-in-Aid for Scientific Research from 
MEXT and JSPS (Nos.\
20028004, 2274016, 22105507).
This work was done in part under the Yukawa International Program for Quark-hadron Sciences (YIPQS).

%\appendix
%\section{First Appendix} %Empty argument \section{} yields `Appendix'. 
%
%\section{Second Appendix}


\begin{thebibliography}{99}
%%%%%%%%%%%%%%%%%%%%%%%%%%%%%%%%%%%%%%%%%%%%%%%%%%%%%%%%%%%%%
% Some macros are available for the bibliography:
%  o for general use
%    \JL : general journals                 \andvol : Vol (Year) Page
%  o for individual journal 
%    \AJ   : Astrophys. J.           \NC         : Nuovo Cim.
%    \ANN  : Ann. of Phys.           \NPA, \NPB  : Nucl. Phys. [A,B]
%    \CMP  : Commun. Math. Phys.     \PLA, \PLB  : Phys. Lett. [A,B]
%    \IJMP : Int. J. Mod. Phys.      \PRA - \PRE : Phys. Rev. [A-E]     
%    \JHEP : J. High Energy Phys.    \PRL        : Phys. Rev. Lett.
%    \JMP  : J. Math. Phys.          \PRP        : Phys. Rep.
%    \JP   : J. of Phys.             \PTP        : Prog. Theor. Phys.     
%    \JPSJ : J. Phys. Soc. Jpn.      \PTPS       : Prog. Theor. Phys. Suppl.
% Usage:
%  \PRD{45,1990,345}          ==> Phys.~Rev.\ D \textbf{45} (1990), 345
%  \JL{Nature,418,2002,123}   ==> Nature \textbf{418} (2002), 123
%  \andvol{123,1995,1020}    ==> \textbf{123} (1995), 1020
%%%%%%%%%%%%%%%%%%%%%%%%%%%%%%%%%%%%%%%%%%%%%%%%%%%%%%%%%%%%%

\bibitem{deeppiexp}
%\cite{Geissel:2002ur}
%\bibitem{Geissel:2002ur}
  H.~Geissel {\it et al.},
  %``Deeply bound 1s and 2p pionic states in Pb-205 and determination of the S
  %wave part of the pion nucleus interaction,''
  Phys.\ Rev.\ Lett.\  {\bf 88}, 122301 (2002);
  %%CITATION = PRLTA,88,122301;%%
%\cite{Itahashi:1999qb}
%\bibitem{Itahashi:1999qb}
  K.~Itahashi {\it et al.},
  %``Deeply Bound Pi- States In Pb-207 Formed In The Pb-208(D, He-3) Reaction
  %Part Ii: Deduced Binding Energies And Widths And The Pion-Nucleus
  %Interaction,''
  Phys.\ Rev.\  C {\bf 62} (2000) 025202.
  %%CITATION = PHRVA,C62,025202;%%


\bibitem{Suzuki:2002ae}
K.~Suzuki {\em et~al.},
\newblock Phys. Rev. Lett. {\bf 92} (2004) 072302;
%\cite{Kienle:2004hq}
%\bibitem{Kienle:2004hq}
P.~Kienle and T.~Yamazaki,
%``Pions In Nuclei, A Probe Of Chiral Symmetry Restoration,''
Prog.\ Part.\ Nucl.\ Phys.\  {\bf 52} (2004) 85.
%%CITATION = PPNPD,52,85;%%

\bibitem{Kolomeitsev:2002gc}
E.~E. Kolomeitsev, N.~Kaiser, and W.~Weise,
\newblock Phys. Rev. Lett. {\bf 90} (2003) 092501.
%%CITATION = NUCL-TH 0207090;%%

%\cite{Jido:2008bk}
\bibitem{Jido:2008bk}
  D.~Jido, T.~Hatsuda and T.~Kunihiro,
  %``In-medium Pion and Partial Restoration of Chiral Symmetry,''
  Phys.\ Lett.\  B {\bf 670} (2008) 109;
  %[arXiv:0805.4453 [nucl-th]].
  %%CITATION = PHLTA,B670,109;%%
%\cite{Jido:2007yt}
%\bibitem{Jido:2007yt}
  %D.~Jido, T.~Hatsuda and T.~Kunihiro,
  %``In-medium Pions and Partial Restoration of Chiral Symmetry: a
  %model-independent analysis,''
  Prog.\ Theor.\ Phys.\ Suppl.\  {\bf 168} (2007) 478.
  %[arXiv:0706.0258 [nucl-th]].
  %%CITATION = PTPSA,168,478;%%

\bibitem{Friedman:2004jh}
E.~Friedman {\em et~al.},
Phys. Rev. Lett. {\bf 93} (2004) 122302;
Phys.\ Rev.\ C {\bf 72} (2005) 034609.

%\cite{Ohnishi:2008ng}
\bibitem{Ohnishi:2008ng}
  A.~Ohnishi, D.~Jido, T.~Sekihara and K.~Tsubakihara,
  %``Possibility of s-wave pion condensates in neutron stars revisited,''
  Phys.\ Rev.\  C {\bf 80} (2009) 038202.
  %[arXiv:0810.3531 [nucl-th]].
  %%CITATION = PHRVA,C80,038202;%%

\bibitem{Glashow:1967rx}
S.~L.~Glashow and S.~Weinberg,
% Breaking chiral symmetry
\newblock Phys. Rev. Lett. {\bf 20} (1968) 224.
%%CITATION = PRLTA,20,224;%%

%\cite{Jido:2000bw}
\bibitem{Jido:2000bw}
  D.~Jido, T.~Hatsuda and T.~Kunihiro,
  %``In-medium pi pi correlation induced by partial restoration of chiral
  %symmetry,''
  Phys.\ Rev.\  D {\bf 63} (2001) 011901(R).
  %[arXiv:hep-ph/0008076].
  %%CITATION = PHRVA,D63,011901;%%

\bibitem{Hyodo:2008xr}
  T.~Hyodo, D.~Jido and A.~Hosaka,
  %``Origin of the resonances in the chiral unitary approach,''
  Phys.\ Rev.\  C {\bf 78} (2008) 025203.
  %[arXiv:0803.2550 [nucl-th]].
  %%CITATION = PHRVA,C78,025203;%%

\bibitem{Jido:2003cb}
  D.~Jido, J.A.~Oller, E.~Oset, A.~Ramos and U.G.~Meissner,
  Nucl.\ Phys.\  A {\bf 725} (2003) 181.
  %%CITATION = NUPHA,A725,181;%%
  
\bibitem{Jido:2009jf}
  D.~Jido, E.~Oset and T.~Sekihara,
  %``Kaonic production of Lambda(1405) off deuteron target in chiral dynamics,''
  Eur.\ Phys.\ J.\  A {\bf 42} (2009) 257.
  %%CITATION = ARXIV:0904.3410;%%
  
\bibitem{Braun:1977wd}
  O.~Braun {\it et al.},
  %``New Information About The Kaon-Nucleon-Hyperon Coupling Constants: G/Anti-K
  %N Sigma (1197), G / Anti-K N Sigma (1385) And G Anti-K N Lambda (1405),''
  Nucl.\ Phys.\  B {\bf 129}, 1 (1977).
  %%CITATION = NUPHA,B129,1;%%

\bibitem{Noumi:JPARC}
  M.~Noumi {\it et al.}, J-PARC proposal E31 
  ``Spectroscopic study of hyperon resonances below $\bar KN$ threshold 
  via the $(K^{-},n)$ reaction on Deuteron'' (2009). 

%\cite{Jido:2010rx}
\bibitem{Jido:2010rx}
  D.~Jido, E.~Oset and T.~Sekihara,
  %``Kaon induced Lambda(1405) production on a deuteron target at DAFNE,''
  arXiv:1008.4423 [nucl-th].
  %%CITATION = ARXIV:1008.4423;%%

\bibitem{nogami}
Y. Nogami,
% Possible existence of kbar nn bound states
Phys. Lett. {\bf 7}, 288 (1963).
  
\bibitem{akaishi02}
  Y.~Akaishi and T.~Yamazaki,
  %``Nuclear anti-K bound states in light nuclei,''
  Phys.\ Rev.\  C {\bf 65}, 044005 (2002).

\bibitem{Ikeda:2007nz}
Y. Ikeda and T. Sato, Phys.\ Rev.\ C {\bf 76}, 035203 (2007);
%\bibitem{ikeda09}
%  Y.~Ikeda and T.~Sato,
  %``On the resonance energy of the barKNN-piYN system,''
  Phys.\ Rev.\  C {\bf 79}, 035201 (2009);
  %[arXiv:0809.1285 [nucl-th]].
  %%CITATION = PHRVA,C79,035201;%%
%\cite{Ikeda:2010tk}
%\bibitem{Ikeda:2010tk}
  Y.~Ikeda, H.~Kamano and T.~Sato,
  %``Energy dependence of barKN interactions and resonance pole of strange
  %dibaryons,''
  arXiv:1004.4877 [nucl-th].
  %%CITATION = ARXIV:1004.4877;%%

\bibitem{shevchenko07}
N.~V.~Shevchenko, A.~Gal, and J.~Mares, Phys.\ Rev.\ Lett.\ {\bf 98}, 
082301 (2007);
%[nucl-th/0610022]
%%CITATION = NUCL-TH/0610022;%%
N. V. Shevchenko, A. Gal, J. Mares and J. Revai, 
Phys. \ Rev.\ C {\bf 76}, 044004 (2007).

\bibitem{dote08}
  A.~Dote, T.~Hyodo and W.~Weise,
  %``$K^-pp$ system with chiral SU(3) effective interaction,''
  Nucl.\ Phys.\  A {\bf 804}, 197 (2008); 
  %arXiv:0806.4917 [nucl-th].
  %%CITATION = ARXIV:0806.4917;%%
%\cite{Dote:2008hw}
%\bibitem{Dote:2008hw}
  %A.~Dote, T.~Hyodo and W.~Weise,
  %``Variational calculation of the ppK^- system based on chiral SU(3)
  %dynamics,''
  Phys.\ Rev.\  C {\bf 79}, 014003 (2009)
  %[arXiv:0806.4917 [nucl-th]].
  %%CITATION = PHRVA,C79,014003;%%

%\cite{Wycech:2008wf}
\bibitem{Wycech:2008wf}
  S.~Wycech and A.~M.~Green,
  %``Variational calculations for K-few-nucleon systems,''
  Phys.\ Rev.\  C {\bf 79}, 014001 (2009).
  %[arXiv:0808.3329 [nucl-th]].
  %%CITATION = PHRVA,C79,014001;%%

\bibitem{Hyodo:2007jq}
T. Hyodo and W. Weise,
% {effective kbar n interaction based on chiral su(3) dynamics}
Phys. Rev. {\bf C77} (2008) 035204.
%[0712.1613]
%%CITATION = 0712.1613;%%

\bibitem{Ikeda:2010}
  Y.~Ikeda, T.~Hyodo, D.~Jido, H.~Kamano, T.~Sato and K.~Yazaki, in preparation;
  A preliminary discussion is found also in 
%\cite{Jido:2010ag}
%\bibitem{Jido:2010ag}
  D.~Jido, T.~Sekihara, Y.~Ikeda, T.~Hyodo, Y.~Kanada-En'yo and E.~Oset,
  %``The nature of Lambda(1405) hyperon resonance in chiral dynamics,''
  Nucl.\ Phys.\  A {\bf 835} (2010) 59.
  %[arXiv:1003.4560 [nucl-th]].
  %%CITATION = NUPHA,A835,59;%%

%\cite{Jido:2008zz}
\bibitem{Jido:2008zz}  
D.~Jido and Y.~Kanada-En'yo,
  %``K anti-K N molecule state with I=1/2 and J^P=1/2^+ studied with three-body
  %calculation,''
  Phys.\ Rev.\  C {\bf 78} (2008) 035203.
  %%CITATION = ARXIV:0806.3601;%%

%\cite{MartinezTorres:2008kh}
\bibitem{MartinezTorres:2008kh}
  A.~Martinez Torres, K.~P.~Khemchandani and E.~Oset,
  %``Solution to Faddeev equations with two-body experimental amplitudes as
  %input and application to J^P=1/2^+, S=0 baryon resonances,''
  Phys.\ Rev.\  C {\bf 79} (2009) 065207.
  %[arXiv:0812.2235 [nucl-th]].
  %%CITATION = PHRVA,C79,065207;%%


%\cite{MartinezTorres:2010zv}
\bibitem{MartinezTorres:2010zv}
  A.~Martinez Torres and D.~Jido,
  %``$K\Lambda(1405)$ configuration of the $K\bar{K}N$ system,''
  Phys.\ Rev.\  C {\bf 82} (2010) 038202.
  %[arXiv:1008.0457 [nucl-th]].
  %%CITATION = PHRVA,C82,038202;%%

%\cite{MartinezTorres:2009cw}
\bibitem{MartinezTorres:2009cw}
  A.~Martinez Torres, K.~P.~Khemchandani, U.~G.~Meissner and E.~Oset,
  %``Searching for signatures around 1920-MeV of a N* state of three hadron
  %nature,''
  Eur.\ Phys.\ J.\  A {\bf 41} (2009) 361.
  %[arXiv:0902.3633 [nucl-th]].
  %%CITATION = EPHJA,A41,361;%%


%%\cite{KanadaEn'yo:2008wm}
%\bibitem{KanadaEn'yo:2008wm}
%  Y.~Kanada-En'yo and D.~Jido,
%  %``anti-K anti-K N molecule state in three-body calculation,''
%  Phys.\ Rev.\  C {\bf 78} (2008) 025212.
  %%CITATION = ARXIV:0804.3124;%%

\bibitem{Haider}
  Q.~Haider and L.~C.~Liu,
  %``Formation Of An Eta Mesic Nucleus,''
  Phys.\ Lett.\  B {\bf 172} (1986) 257,
  %%CITATION = PHLTA,B172,257;%%
  L.~C.~Liu and Q.~Haider,
  %``Signature For The Existence Of Eta Mesic Nucleus,''
  Phys.\ Rev.\  C {\bf 34} (1986) 1845.
  %%CITATION = PHRVA,C34,1845;%%

\bibitem{Jido:2002yb}
D.~Jido, H.~Nagahiro, and S.~Hirenzaki, Phys. Rev. C {\bf 66}  (2002) 045202.
%%CITATION = NUCL-TH/0206043;%%
  
\bibitem{DK}
C.~DeTar and T.~Kunihiro, Phys. Rev. D {\bf 39} (1989) 2805.

\bibitem{cdmod}
D.~Jido, Y.~Nemoto, M.~Oka, and A.~Hosaka, Nucl. Phys. A {\bf 671} (2000) 471;
D.~Jido, M.~Oka, and A.~Hosaka, Prog. Theor. Phys. {\bf 106} (2001) 873.
%%CITATION = HEP-PH/9805306;%%
%%CITATION = HEP-PH/0110005;%%

%\cite{Jido:2008ng}
\bibitem{Jido:2008ng}
  D.~Jido, E.~E.~Kolomeitsev, H.~Nagahiro and S.~Hirenzaki,
  %``Level crossing of particle-hole and mesonic modes in eta mesic nuclei,''
  Nucl.\ Phys.\ A {\bf 811}, 158 (2008).
%  arXiv:0801.4834 [nucl-th].
  %%CITATION = ARXIV:0801.4834;%%

\bibitem{Hayano:1998sy}
R. S.~Hayano, S.~Hirenzaki, and A.~Gillitzer,
% Formation of eta mesic nuclei using the recoilless (d,he-3) reaction
Eur. Phys. J. A {\bf 6} (1999) 99.
%[nucl-th/9806012]
%%CITATION = NUCL-TH/9806012;%%

\bibitem{Nagahiro:2003iv}
H.~Nagahiro, talk in this conference;
H.~Nagahiro, D.~Jido, and S.~Hirenzaki, Phys.\ Rev.\ C {\bf 68} (2003) 035205;
Nucl.\ Phys.\ A {\bf 761}, 92 (2005);
%%CITATION = NUCL-TH/0304068;%%
%\cite{Nagahiro:2008rj}
%\bibitem{Nagahiro:2008rj}
%  H.~Nagahiro, D.~Jido and S.~Hirenzaki, 
  Phys.\ Rev.\ C {\bf 80} (2009) 025205.
  %``Formation of eta-mesic nuclei by (pi,N) reaction and N^*(1535) in medium,''
  %arXiv:0811.4516 [nucl-th].
  %%CITATION = ARXIV:0811.4516;%%


\end{thebibliography}
\end{document}